\documentclass[preprint,showpacs,preprintnumbers,amsmath,amssymb]{revtex4}

\usepackage{graphicx}
\usepackage{dcolumn}
\usepackage{bm}
\usepackage{epsfig}

\begin{document}

\title{Magnetism and quantum phase transitions in spin-1/2
  attractive fermions with polarization}
\author{J.-S. He$^{1}$, A. Foerster$^{2}$, X. W. Guan$^{3}$ and 
M. T. Batchelor$^{3,4}$}
\address{
$^{1}$ Department of Mathematics, Ningbo University, Ningbo,
315211 Zhejiang, P.R. China\\
$^{2}$ Instituto de Fisica da UFRGS,  Av.  Bento Goncalves, 9500,
                     Porto Alegre, 91501-970, Brasil\\
$^{3}$ Department of
Theoretical Physics, Research School of Physics and
Engineering, Australian National University, Canberra ACT 0200,
Australia\\
$^{4}$ Mathematical Sciences Institute,
Australian National University, Canberra ACT 0200,
Australia}

\date{\today}

\begin{abstract}
\noindent
An extensive investigation is given for magnetic properties and phase transitions
in one-dimensional Bethe ansatz integrable spin-1/2 attractive fermions with polarization
by means of the dressed energy formalism.  An iteration method is presented to 
derive higher order corrections for the ground state energy, critical fields and magnetic
properties. Numerical solutions of the dressed energy equations 
confirm that the analytic expressions for these physical quantities and resulting phase diagrams 
are highly accurate in the weak and strong coupling regimes, capturing
the precise nature of magnetic effects and quantum phase transitions in one-dimensional
interacting fermions with population imbalance. Moreover, it is shown that the
universality class of linear field-dependent behaviour of the
magnetization holds throughout the whole attractive regime. 

\end{abstract}



\maketitle

\section{Introduction}

Bosons and fermions reveal strikingly different quantum statistical effects at low temperatures.
Bosons with integer spin undergo Bose-Einstein condensation (BEC), whereas
fermions with half-odd-integer spin are not allowed to occupy a single quantum state 
due to the Pauli exclusion principle. 
However, fermions with opposite spin states can
pair up to produce Bardeen-Cooper-Schrieffer (BCS) pairs to
form a Fermi superfluid.  
Quantum degenerate gases of ultracold atoms
open up exciting possibilities for the experimental study of such subtle
quantum many-body physics in low dimensions
\cite{1D-1,Lewenstein,Grimm,Ketterle}. 
In this platform, Feshbach
resonance has given rise to a rich avenue for the experimental
investigation of relevant problems, such as the crossover from BCS
superfluidity to BEC \cite{BCS-BEC}, fermionic superfluidity and phase
transitions, among others \cite{I-F1,I-F2,I-F3}. 
Particularly, pairing
and superfluidity are attracting further attention from theory and
experiment due to the close connection to high-T$_c$ superconductivity
and nuclear physics.  The study of pairing signature and fermionic
superfluidity in interacting fermions has stimulated growing interest
in Fermi gases with population imbalance \cite{im1,im2,im3,im4}, i.e.,
systems with different species of fermions \cite{Grimm} as well as
multicomponent interacting fermions
\cite{Rapp,Lecheminant,Demler,Wilczek}.  This gives rise to new
perspectives to explore subtle quantum phases, such as a breached
pairing phase \cite{Sama} and a nonzero momentum pairing phase of
Fulde-Ferrell-Larkin-Ovchinnikov (FFLO) states \cite{FFLO,MF-FFLO} and 
colour superfluids \cite{Ketterle}.

One-dimensional (1D) atomic gases with internal
degrees of freedom also provide tunable interacting many-body systems featuring
novel magnetic properties and quantum phase transitions
\cite{Gia,Cazalilla1,Fuchs,BBGO,Giaradeau}. Although the FFLO state
has not been fully confirmed experimentally, investigations of
the elusive FFLO state in the 1D interacting Fermi gas with population
imbalance are very promising
\cite{Orso,Hu,GBLB,Feiguin,Wadati,Ueda,Batrouni,Parish,Ying,cazalilla2,Liu,Mueller,Chen,Luscher,Tsvelik,Bolech}.
Theoretical predictions for the existence of a FFLO state in the 1D
interacting Fermi gas has emerged by a variety of methods including
the Bethe ansatz (BA) solution \cite{Hu}, numerical methods
\cite{Feiguin,Ueda,Batrouni,Parish,cazalilla2,Mueller,Bolech} and field
theory \cite{Liu,Yang-K,Vekua}. A powerful field theory approach
\cite{Liu,Vekua} was used to describe a FFLO state in the 1D Fermi
superliquid with population imbalance. Nevertheless, 
verification of the FFLO signature of polarized attractive
fermions is still lacking via the Bethe ansatz solution.  A recent
thermodynamic Bethe ansatz (TBA) study of strongly attractive
fermions \cite{GBLB} shows that paired and unpaired atoms form
two Fermi liquids coupled to each other. The TBA equations indicate
that spin wave fluctuations ferromagnetically couple to the unpaired
Fermi sea. A full analysis of magnetic effects and low energy physics of 
spin-1/2 fermions with polarization in both the weak and strong coupling
regimes, as well as a detailed discussion on the universality class
of the magnetic behaviour in the whole attractive regime, are
desirable in understanding such subtle paired states in 1D
interacting fermions with polarization.  

In this paper we provide an extended investigation of quantum phases
and phase transitions for 1D interacting fermions with
polarization in the presence of an external field. We analytically and
numerically solve the dressed energy equations which describe the
equilibrium state at zero temperature. 
We extend previous work on this model to derive higher order
corrections (up to order $1/|\gamma|^3$) for the ground state
energy, magnetization, critical fields, chemical potentials and the
external field-energy transfer relation.
The phase diagrams in the weak and strong
coupling regimes are obtained in terms of the external field, density
and interaction strength.  In the strong coupling regime, (i) the bound
pairs in the homogeneous system form a singlet ground state when the
external field is less than the lower critical value $H_{c1}$, (ii) a
normal Fermi liquid phase without pairing occurs when the external
field is greater than the upper critical value $H_{c2}$ and (iii) for an
intermediate range $H_{c1} < H < H_{c2}$, paired and unpaired
atoms coexist.  
However, for weak coupling, a BCS-like pair scattering
phase occurs only when the external field $H =0$, while paired and
unpaired fermions coexist when the field is less than a critical
field. Significantly, we also show that the universality class of
linear field-dependent behaviour of the magnetization remains throughout the whole
attractive regime.

This paper is set out as follows. In section \ref{model}, we present
the Hamiltonian and discuss the pairing signature for the 1D fermions
with population imbalance in the whole attractive regime. In section
\ref{TBA}, we present the dressed energy equations obtained from the
thermodynamic Bethe ansatz equations (TBA) in the limit $T\to 0$. In
section \ref{magnetization}, we present the magnetic properties for
the model in the weak coupling regime. We solve the dressed energy
equations in the strong coupling regime in section \ref{DEE}. The
explicit forms of the magnetic properties and the ground state energy are
given in terms of the interaction strength, density and external
field.  In section \ref{diagram}, we present the full phase diagrams for
the whole attractive regime.  Section \ref{Conclusion} is devoted to
concluding remarks and a brief discussion.

\section{The model}
\label{model}

The Hamiltonian  \cite{Yang,Gaudin}  we consider
\begin{eqnarray}
{\cal H} &=& \sum _{j=\downarrow,\uparrow} \int_0^L \phi _{j}^{\dagger}(x) \left
(-\frac{\hbar^{2}}{2m}\frac{d^{2}}{dx^{2}}  \right ) \phi
_{j}^{}(x) dx\nonumber\\
&& +  g_{1D} \int_0^L  \phi _{\downarrow}^{\dagger}(x) \phi
_{\uparrow}^{\dagger}(x) \phi _{\uparrow}^{}(x)
\phi _{\downarrow}^{}(x) dx \nonumber\\
&& - \frac12{H}\int_0^L  \left (\phi _{\uparrow}^{\dagger}(x) \phi
_{\uparrow}^{}(x) - \phi _{\downarrow}^{\dagger}(x) \phi
_{\downarrow}^{}(x) \right ) dx. \label{Ham}
\end{eqnarray}
describes $N$ $\delta$-interacting spin-$\frac{1}{2}$ fermions 
of mass $m$ constrained by periodic boundary conditions to a line of length $L$ 
and subject to an external magnetic field $H$.  
In this formulation the field operators $\phi_{\downarrow}$ and $\phi_{\uparrow}$ describe 
fermionic atoms in the respective states $| \! \uparrow \rangle$ and $| \! \downarrow \rangle$. 
The $\delta$-type interaction between 
fermions with opposite hyperfine states preserves the spin states such
that the Zeeman term in the Hamiltonian (\ref{Ham}) is a conserved quantity.  
For convenience, we use units of $\hbar =2m=1$ and define
$c = mg_{1D}/\hbar ^2$ and a dimensionless interaction strength
$\gamma=c/n$ for the physical analysis, where $n = {N}/{L}$ is the linear density.
The inter-component interaction can be tuned from strongly attractive
($g_{\rm 1D}\rightarrow -\infty$) to strongly repulsive ($g_{\rm 1D}
\rightarrow +\infty$) via Feshbach resonance and optical
confinement. The interaction is attractive for $g_{\rm 1D}<0$ and
repulsive for $g_{\rm 1D}>0$.

The model (\ref{Ham}) was solved by Yang \cite{Yang} and Gaudin \cite{Gaudin}
in the 1960's and has received renewed interest in connection with 
ultracold atomic gases
\cite{Gia,Cazalilla1,Fuchs,BBGO,Giaradeau,Orso,Hu,GBLB,Feiguin,Wadati,Ueda,Batrouni,Parish,cazalilla2,Liu,Mueller,Chen,Luscher,Tsvelik,Bolech}. 
The energy eigenspectrum is given by 
$E=\frac{\hbar ^2}{2m}\sum_{j=1}^Nk_j^2$, where the Bethe ansatz (BA) wave numbers
$\left\{k_i\right\}$ and the rapidities $\left\{\Lambda_{\alpha}\right\}$ for the internal spin degrees
of freedom satisfy the BA equations
\begin{eqnarray}
\exp(\mathrm{i}k_jL)&=&\prod^M_{\ell = 1}
\frac{k_j-\Lambda_\ell+\mathrm{i}\, c/2}{k_j-\Lambda_\ell-\mathrm{i}\, c/2} \nonumber\\
\prod^N_{\ell = 1}\frac{\Lambda_{\alpha}-k_{\ell}+\mathrm{i}\, c/2}{\Lambda_{\alpha}-k_{\ell}-\mathrm{i}\, c/2}
 &=& - {\prod^M_{ \beta = 1} }
\frac{\Lambda_{\alpha}-\Lambda_{\beta} +\mathrm{i}\, c}{\Lambda_{\alpha}-\Lambda_{\beta} -\mathrm{i}\, c} .
\label{BE}
\end{eqnarray}
Here $j = 1,\ldots, N$ and $\alpha = 1,\ldots, M$, with $M$ the number
of spin-down fermions.

\begin{center}
\begin{figure}
\includegraphics[width=0.9\linewidth]{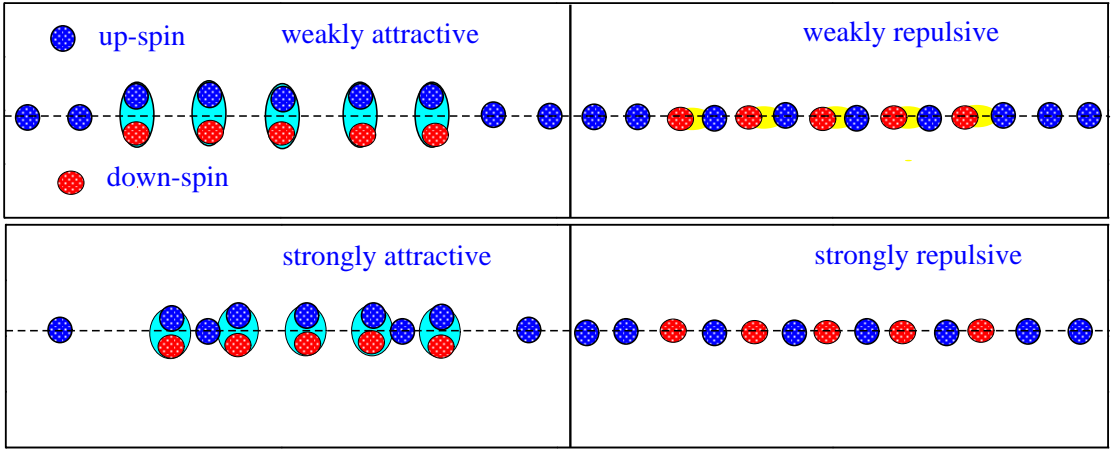}
\caption{Bethe ansatz root configurations for pairing and depairing in
  quasimomentum space.  For weakly attractive interaction, unpaired
  roots sit in the out-wings due to strong Fermi pressure. For
  strongly attractive interaction, unpaired roots can penetrate into
  the central region, occupied by the bound pairs. However, for weakly
  repulsive interaction the roots with up- and down- spins separate
  gradually. In the strongly repulsive regime, the model forms an effective
  Heisenberg spin chain with the antiferromagnetic coupling constant
  $J\approx -4E_F/\gamma$ \cite{GBL}, where $E_F=n^2\pi^2/3$ is the Fermi
  energy.}
\label{fig:Cartoon}
\end{figure}
\end{center}

The solutions to the BA equations (\ref{BE}), as depicted in Figure \ref{fig:Cartoon}, provide
a clear pairing signature and the ground state properties of the model.
The BA root distributions in the complex plane were studied recently \cite{Wadati,BBGO}. 
The ground state for 1D interacting fermions with repulsive
interaction has antiferromagnetic ordering
\cite{LiebMattis,Giamarchi,BBGO}.
Rather subtle magnetism for the model with repulsive interaction was recently studied \cite{MC,GBL}. 
For attractive interaction, fermions with different spin states can
form BCS pairs with nonzero centre-of-mass momenta which
might feature FFLO states \cite{Hu,Feiguin,Ueda,Batrouni,Parish,cazalilla2,Liu}. %

In the weakly attractive regime, the weakly bound Cooper pairs are
not stable due to thermal and spin wave fluctuations. The unpaired
fermions sit on two outer wings in the quasimomentum space
\cite{BBGO,GBLB} due to the Fermi pressure (see Figure \ref{fig:Cartoon}). 
The ground state can only have one pair of fermions with opposite
spins having a particular quasimomentum $k$. The paired fermions
occupy the central area in the quasimomentum $k$ space.  Indeed we
find from the BA equations (\ref{BE}) that in the weak coupling limit, i.e. $L|c|\ll 1$,
the imaginary part of the quasimomenta for a BCS pair is proportional
to $\sqrt{|c|/L}$. However, for strongly attractive interaction,
i.e. $L|c| \gg 1$, the BCS pair has imaginary part $\pm
 \mathrm{i} \frac12{|c|}$. In this regime, the lowest spin excitation has
an energy gap which is proportional to $c^2$.  For the cross-over
regime, i.e. $L|c| \sim 1$, the imaginary part $\pm \mathrm{i}y$ is
asymptotically determined by the condition $y\tanh (\frac12{yL})\approx \frac12{|c|}$. 
For this cross-over regime, the
spin gap might be exponentially small. However, it is hard to
analytically determine this small energy gap from the BA equations
(\ref{BE}). Nevertheless, for the weak coupling regime $L|c|\ll 1$, the
bound state has a small binding energy 
$\epsilon_{\rm b}={\hbar^2n|\gamma|}/{mL}$ which has the same order of $\gamma$ as
the interacting energies of pair-pair and pair-unpaired fermions.  In
this limit, the real parts of the quasimomenta satisfy the Gaudin
model-like BA equations \cite{BBGO,Gaudin-BCS} which describe BCS pair-pair
and pair-unpaired fermion scattering. They form a gapless
superconducting phase.  Using the above BA root configuration, the ground state
energy per unit length is given by \cite{BBGO}
\begin{equation}
E \approx
\frac{\hbar^2n^3}{2m}\left(-\frac{|\gamma|}{2}(1-P^2)+\frac{\pi^2}{12}+\frac{\pi^2}{4}P^2\right) \label{E-w}
\end{equation}
in terms of the polarization $P=(N-2M)/N$. The first term in
Eq. (\ref{E-w}) includes the collective interaction energy 
(pair-pair and pair-unpaired fermion scattering energy) and the
binding energy (internal energy). We see clearly that for large
polarization ($P \approx 1)$ the small portion of spin-down fermions are 
likely to experience a mean-field
formed by the spin-up medium. This is consistent with the observation of
Fermi polarons in an attractive Fermi liquid of ultracold atoms \cite{MIT}.

On the other hand, when the attractive interaction strength is
increased, i.e. $L|c|\gg 1$, the bound pairs gradually form hard-core
bosons while the unpaired fermions can penetrate into the central
region in the quasimomentum space (see Figure \ref{fig:Cartoon}).
The main reason for the unpaired fermions and BCS pairs having
overlapping Fermi seas is that in 1D the paired and unpaired fermions
have different fractional statistical signatures such that they are
allowed to pass into each other in the quasimomentum space.
In the thermodynamic limit, i.e., $L\to \infty$, $N \to \infty$ with $N/L$ finite, the
binding energy of a pair is $\epsilon_{\rm b}= {\hbar^2n^2\gamma^2}/{(4m)}$. 
The dimensionless interaction strength $\gamma=c/n$ is inversely
proportional to the density $n$. 
This signature leads to different phase segments in 1D trapped fermions
\cite{Orso,Hu,GBLB} than the phase separations in 3D trapped
interacting fermions, where the Fermi gas has been separated into a
uniformly paired inner core surrounded by a shell with the excess of
unpaired atoms \cite{I-F1,I-F2,I-F3}.

From the ground state energy for the model with strong
attraction and arbitrary polarization \cite{BBGO}, we find 
the finite-size corrections to the energy 
in the thermodynamic limit to be given by
\begin{equation}
E(L,N)-LE_0^{\infty} \approx -\frac{\pi \hbar C}{6L}(v_{\rm
b}+v_{\rm u}),
\end{equation}
where the central charge $C=1$ and the group velocities for
bound pairs $v_b$ and unpaired fermions $v_u$ are 
\begin{eqnarray}
v_{\rm b} &\approx &\frac{v_{\rm F}(1-P)}{4}
\left(1+\frac{(1-P)}{|\gamma|} +\frac{4P}{|\gamma|}\right) \nonumber\\
v_{\rm u}&\approx & v_{\rm F} P\left( 1 +
\frac{4(1-P)}{|\gamma|}\right).
\end{eqnarray}
Here the Fermi velocity is $v_{\rm F}=\hbar \pi n/m$.
In the above equation $E_0^{\infty}$ is the ground state energy in the 
thermodynamic limit
\begin{eqnarray}
E_0^{\infty} &\approx  & \frac{\hbar^2n^3}{2m} \left\{-\frac{(1-P)\gamma^2}{4} +
\frac{P^3\pi^2}{3} \left(1+\frac{4(1-P)}{|\gamma|}\right)\right.\nonumber\\
& & \left. + \frac{\pi^2(1-P)^3}{48}
\left(1+\frac{(1-P)}{|\gamma|}+\frac{4P}{|\gamma|}\right)\right\} .
\label{E-P}
\end{eqnarray}
The nature of the finite-size corrections indicate that two Fermi liquids couple to each
other and have different statistical signatures. The low energy
physics is dominated by the charge density fluctuations. 
The spin wave fluctuations are frozen out.  
In order to understand the pairing signature and the subtle FFLO
states in 1D, one should investigate density distributions, pairing
correlations and thermodynamics, which we do here through the
thermodynamic Bethe ansatz formalism.
In particular, we shall focus on magnetic properties and
quantum phase transitions for the whole attractive regime.

\section{Thermodynamic Bethe Ansatz}
\label{TBA}

The thermodynamic Bethe ansatz (TBA) provides a powerful and elegant way to study the 
thermal properties of 1D integrable systems. 
It also provides a convenient formalism  to analyze quantum phase transitions and magnetic effects in the presence
of external fields at zero temperature \cite{Takahashi,Hubbardbook,Schlot,BGOT}.  In the
thermodynamic limit, the grand partition function is
$Z={\mathrm {tr}} (\mathrm{e}^{-\cal{H}/T})=\mathrm{e}^{-G/T}$, in terms of the Gibbs
free energy $G = E - HM^z - \mu n - TS$ and the
magnetization $H$, the chemical potential $\mu$ and the entropy $S$
\cite{Takahashi,Hubbardbook,Schlot,BGOT}. 
The TBA equations for the attractive
regime are much more subtle and involved compared to those for the repulsive regime. 
In general the equilibrium states satisfy the condition of
minimizing the Gibbs free energy $G$ with respect to the particle and
hole densities for the charge and spin degrees of freedom that generates the
TBA equations (details are given in \cite{Lai-1,Takahashi,BGOT}).  
At zero temperature, the ground state properties
are determined in terms of the dressed energies for the paired $\epsilon^{\rm b}$
and unpaired fermions $\epsilon^{\rm u}$ and the function
\begin{equation}
a_m(x)=\frac{1}{2\pi}\frac{m|c|}{(m \,c/2)^2+x^2}
\end{equation}
by 
\begin{eqnarray}
\epsilon^{\rm b}(\Lambda)&=&2(\Lambda^2-\mu
-\frac14{c^2})-\int_{-B}^{B}a_2(\Lambda-\Lambda'){\epsilon^{\rm
    b}}^{-}(\Lambda')d\Lambda' \nonumber\\
& &-\int_{-Q}^{Q}a_1(\Lambda-k){\epsilon^{\rm u}}^{-}(k)d k,\nonumber\\
\epsilon^{\rm u}(k)&=&(k^2-\mu
-\frac12{H})-\int_{-B}^{B}a_1(k-\Lambda){\epsilon^{\rm
    b}}^{-}(\Lambda)d\Lambda \label{TBA-F}
\end{eqnarray}
which are the dressed energy equations \cite{Takahashi,Schlot,GBLB}
obtained from the TBA equations in the limit $T\to 0$. 
The superscripts $\pm$ denote the
positive and negative parts of the dressed energies, with the negative (positive) 
part corresponding to occupied (unoccupied) states.
The integration boundaries $B$ and $Q$ characterize the Fermi surfaces
of the bound pairs and unpaired fermions, respectively.  

The Gibbs free energy per unit length at zero temperature is given by
\begin{equation}
G(\mu,H)=\frac{1}{\pi}\int_{-B}^B{\epsilon^{\rm b}}^-(\Lambda
)d\Lambda+\frac{1}{2\pi}\int_{-Q}^Q{\epsilon^{\rm u}}^-(k)dk.
\end{equation}
The TBA equations provide
a clear configuration for band fillings with respect to the external
field $H$ and chemical potential $\mu$.  The polarization $P$ varies
with respect to the external magnetic field.  From the Gibbs free energy
we have the relations
\begin{equation}
 -\partial G(\mu,H)/\partial \mu =n, \,\,\,\,-\partial
G(\mu,H)/\partial H =n P/2.
\end{equation}

\section{Magnetic properties in the weak coupling regime}
\label{magnetization}

For weak coupling $|c|\to 0$ caution needs to be taken in the thermodynamic limit. 
On solving the discrete BA equations (\ref{BE}) in the regime 
$L|c|\ll 1$ the imaginary part of the BCS-like pairs tends to
$\sqrt{|c|/L}$ \cite{BBGO}.
However, the TBA equations \cite{Takahashi,GBLB} usually follow from
the root patterns in the thermodynamic limit, i.e. $L,N \to \infty$ with $N/L$ is
fixed. Under this limit, we naturally have the BA root patterns 
$k_j=\Lambda_j \pm \mathrm{i}\frac12 |c|$ with $j=1,\ldots, M$ for the
charge degree and the string patterns with equally spaced imaginary
distribution for spin rapidity
$\Lambda_{\alpha,j}^{n}=\Lambda^{(n)}_{\alpha}+\mathrm{i} \frac{1}{2}
(n+1-2j)c$, with $j=1,\ldots ,n$.
Here the number of strings $\alpha=1,\ldots, N_n$.
$\Lambda^n_{\alpha}$ is the position of the centre for the length-$n$
 string on the real axis in $\Lambda$-space.  Therefore, in the weak
 coupling limit, i.e., $|c|\to 0$, the integral BA equations and the
 TBA equations do not properly described the true solutions to the
 discrete BA equations (\ref{BE}) unless under the thermodynamic limit
 conditions.  Nevertheless, the discrepancy is minimal, i.e., it is
 $O(\gamma^2)$.  

The BA equations (\ref{BE}) in principle give complete states of the model.
 However, at finite temperatures, the true physical states become
 degenerate. The dressed energies in the TBA  equations (\ref{TBA-F}) characterize
 excitation energies above the Fermi surfaces of the bound pairs and
 unpaired fermions. All physical quantities, for example, free energy,
 pressure and magnetic properties can be obtained from the TBA
 equations withought deriving the spectral properties of low-lying
 excitations. In the weak coupling limit, the interaction energy is proportional to
 $|c|$ which is much less than the kinetic energy. Therefore, in this
 regime, the exact ground state energy with leading term of order
 $|c|$ is precise enough to capture the nature of phase transitions and magnetic
 ordering.

\begin{figure}[ht]
{{\includegraphics [width=0.60\linewidth,angle=-90]{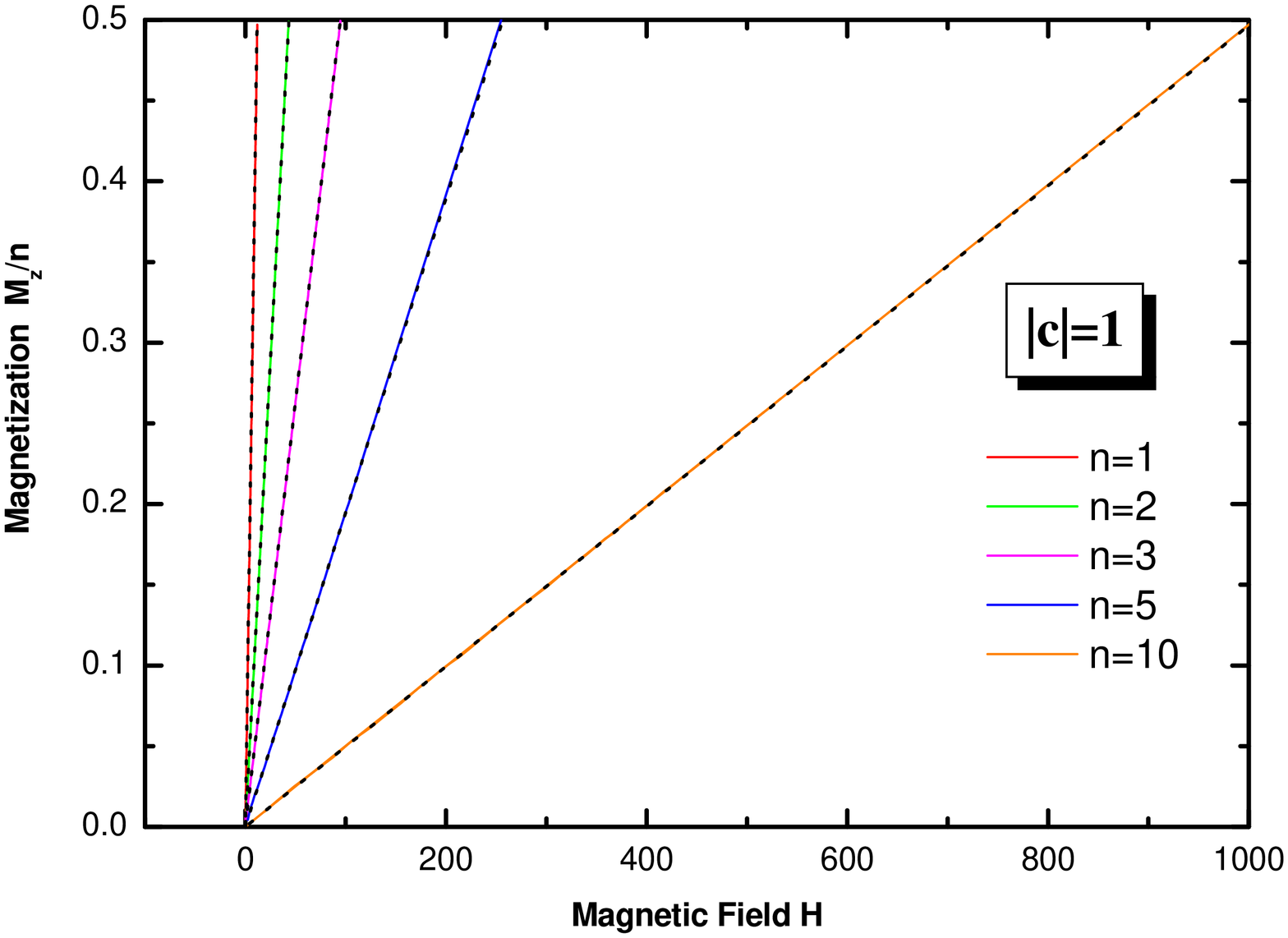}}}
\caption{Magnetization $M^z/n$ vs external field $H$ in units $2m=\hbar=1$ with weak coupling $|c|=1$ 
for different densities $n$. The dashed lines are plotted from the analytic result (\ref{mz-w}).
Excellent agreement between the analytic result and numerical solution of the 
integral equations (\ref{TBA-F}) (solid lines) is seen.}\label{fig:mz-w}
\end{figure}

\begin{figure}[ht]
{{\includegraphics [width=0.60\linewidth,angle=-90]{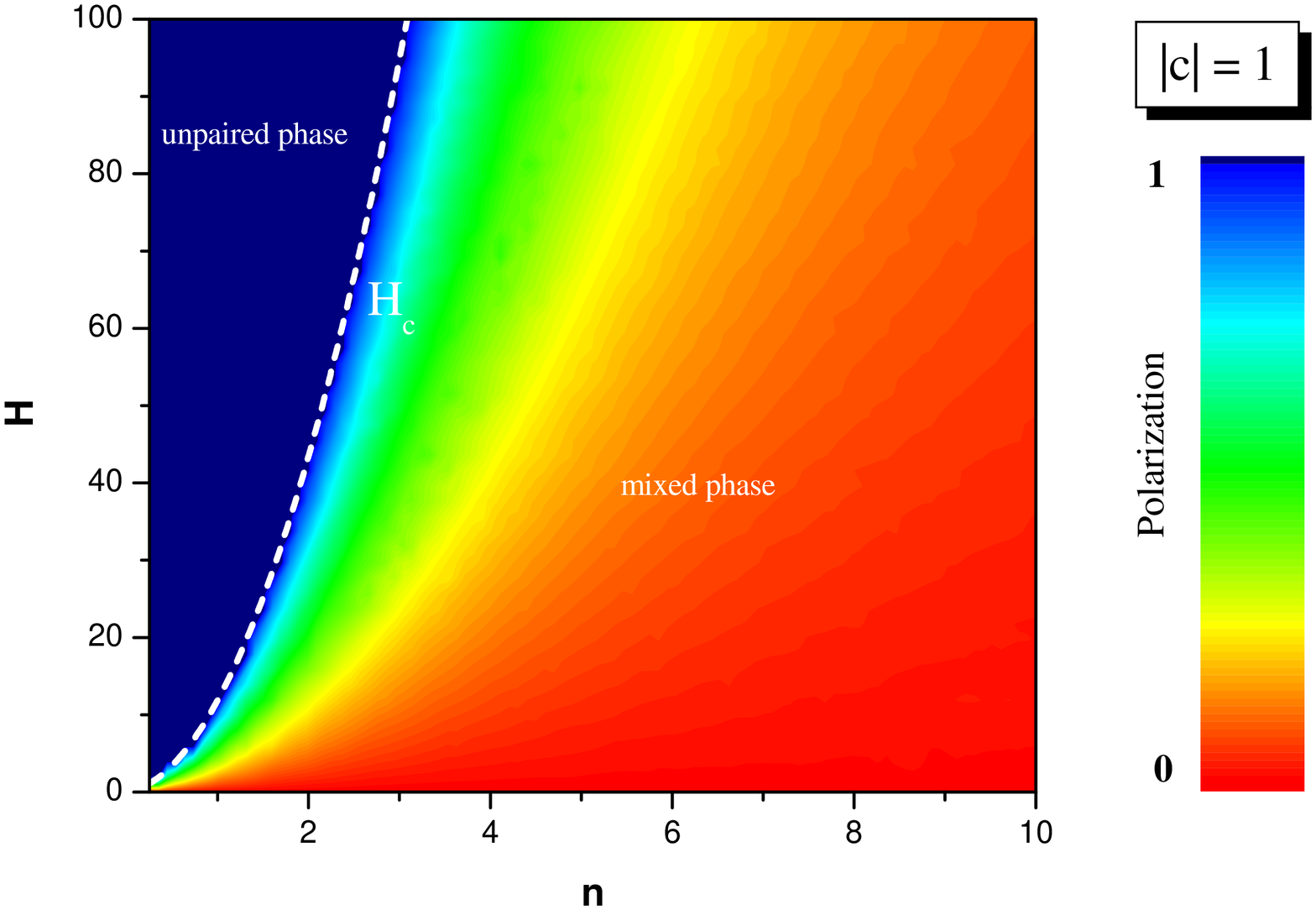}}}
\caption{Phase diagram for weak coupling value $|c| =1$. The dashed
line plotted from the analytic result (\ref{Hc-w}) is in excellent 
agreement with the coloured phases which are obtained from numerical solutions of 
the dressed energy equations (\ref{TBA-F}).  }\label{fig:phase-w}
\end{figure}

From the ground state energy (\ref{E-w}) we have the relation between the
external field and magnetization
\begin{equation}
H\approx \frac{\hbar^2n^2}{2m}\left[2\pi^2 m^z+4|\gamma| m^z\right] \label{mz-w}
\end{equation}
where $m^z={M^z}/{n}$ and the magnetization is defined by
$M^z={nP}/{2}$.  
A linear field-dependent behaviour of the magnetization is observed.  
Figure \ref{fig:mz-w} shows the magnetization vs the field $H$ for different interaction values
$|\gamma|$. 
We observe that the analytic results plotted from (\ref{mz-w})
are in excellent agreement with the numerical curves evaluated directly from the
dressed energy equations (\ref{TBA-F}).  
We also find that a fully paired ground state only occurs in the absence of the external field. 
However, for $H \ge H_{c}$ where
\begin{equation}
H_{c} = n^2[\pi^2+2|\gamma| ]\label{Hc-w}
\end{equation}
the fully polarized phase occurs.  
Paired and unpaired fermions coexist in the intermediate range $0<H<H_{c}$.  
The phase diagram for weak coupling is illustrated in Figure \ref{fig:phase-w}.

\section{Solutions to the dressed energy equations}
\label{DEE}

In this section we solve the dressed energy equations (\ref{TBA-F}) 
analytically in the strong coupling regime to obtain explicit forms for the
critical fields and magnetic properties in terms of the interaction strength $\gamma$. 
We present a systematic way to obtain these physical properties 
up to order $\frac{1}{|\gamma|^3}$ which  gives a 
very precise phase diagram for finite strong interaction. 
Here we note that Iida and Wadati \cite{Wadati} have presented a different 
method to solve the dressed energy equations. 
We have solved the dressed energy equations (\ref{TBA-F}) numerically in the whole attractive regime 
to compare with the analytic results.
Excellent agreement between numerical and analytical results is found.

First consider the ground state $P=0$. 
Following the method developed in \cite{GBLB} where the dressed energy
equations (\ref{TBA-F}) are asymptotically expanded in terms of $1/|c|$, the
ground state dressed energy equation for $P=0$ is given by
\begin{equation}
\epsilon^b(\Lambda)\approx 2(\Lambda^2-\bar\mu )-\frac{|c|}{\pi} \int^B_{-B}
\frac{\epsilon^b(\Lambda')}{c^2+(\Lambda-\Lambda')^2}d\Lambda'
\label{E-B-1}
\end{equation}
with $\bar\mu=\mu+\frac{c^2}{4}$. For convenience, we introduce the notation
\begin{eqnarray}
p^b&=& -\frac{1}{\pi}\int^B_{-B}\epsilon^b(\Lambda)d \Lambda \label{defpb}\\
p^u&=&-\frac{1}{2\pi}\int^Q_{-Q}\epsilon^u(k)dk \label{defpu}
\end{eqnarray}
for the pressure of bound pairs and un-paired fermions.  
Since the Fermi point $B$ is finite, we can take an
expansion with respect to $\Lambda '$ in the integral in Eq. (\ref{E-B-1}). 
By a straightforward calculation, the pressure $p^b$ is found to be
\begin{equation}
p^b\approx-\frac{4}{\pi}\left(  \frac{B^3}{3}-\bar{\mu}B
\right)-\frac{2B}{\pi}\frac{p^b}{|c|}+ \frac{128}{45 \pi^2
|c|^3}(\bar{\mu})^3.\label{pb}
\end{equation}
We obtained this equation by iteration in terms of $p^b$ and
$\bar{\mu}$. In such a way, the accuracy of physical quantities can be
controlled to order $1/|c|$. This provides a systematic way to
obtain accurate results from dressed energy equations. It is free
from restriction on the integration boundaries $B$ and $Q$. 
Furthermore, from  Eq. (\ref{pb}) and the condition $\epsilon^b(\pm B)=0$ we find
\begin{equation}\label{eqBsquarePzero}
B^2\approx\bar{\mu}-\frac{p^b}{2|c|}+\frac{8}{5\pi
|c|^3}(\bar{\mu})^{\frac{5}{2}}.
\end{equation}
Finally, using the above equations and the
relation $\partial p^b/\partial\mu=n$, the pressure per unit length follows as
\begin{eqnarray}
p^b&\approx&\frac{\hbar^2n^3}{2m}\frac{\pi^2n^3}{24}\left(1+\frac{3}{2|\gamma|}
+\frac{3}{2|\gamma|^2}+\frac{1}{4|\gamma|^3}(5-\frac{\pi^2}{3}) \right)
\label{solutionpbPzero}
\end{eqnarray}
and the energy per unit length is 
\begin{eqnarray}
E_0\approx \frac{\hbar^2n^3}{2m}\left\{-\frac{\gamma^2}{4}+\frac{\pi^2}{48}\left[1
+\frac{1}{|\gamma|}+\frac{3}{4|\gamma|^2}+\frac{1}{2|\gamma|^3}\left(1-\frac{\pi^2}{15}\right)\right]\right\}.\label{E0Pzero}
\end{eqnarray}

The dressed energy equations (\ref{TBA-F}) can also be solved analytically for $0<P<1$.
Following \cite{GBLB}, we define $\tilde{\mu}=\mu+ {H}/{2}$. 
We notice that the Fermi points
$Q$ and $B$ are still finite in the presence of an external field $H$. 
Similar to the case $P=0$, using the conditions $\epsilon^b(\pm B)=0$ and $\epsilon^u(\pm Q)=0$, 
we obtain the relations
\begin{eqnarray}
p^b&\approx&-\frac{4}{\pi}\left(\frac{B^3}{3}-\bar\mu
B\right)-\frac{2B}{\pi}\frac{p^b}{|c|}
-\frac{8B}{\pi}\frac{p^u}{|c|}\nonumber\\
& &+\frac{128\bar{\mu}}{45\pi^2|c|^3}+\frac{64(\bar\mu)^{\frac{3}{2}} (\tilde\mu)^{\frac{3}{2}}   }{9\pi^2|c|^3}
+\frac{64(\bar\mu)^{\frac{1}{2}} (\tilde\mu)^{\frac{5}{2}}   }{15\pi^2|c|^3},\\
p^u&\approx&-\frac{Q}{\pi}\left(\frac{Q^2}{3}-\tilde\mu
\right)-\frac{2Q}{\pi}\frac{p^b}{|c|}\nonumber\\
& &+
\frac{64(\bar\mu)^{\frac{3}{2}} (\tilde\mu)^{\frac{3}{2}}   }{9\pi^2|c|^3}
+\frac{64(\tilde\mu)^{\frac{1}{2}} (\bar\mu)^{\frac{5}{2}}
}{15\pi^2|c|^3}
\end{eqnarray}
and
\begin{eqnarray}
B^2 &\approx&
  \bar\mu+\frac{p^b}{2|c|}-\frac{2p^u}{|c|}+\frac{4\bar{\mu}^{\frac{5}{2}}}{3\pi
  |c|^3} \nonumber\\ & &+\frac{16\tilde{\mu}
  ^{\frac{2}{2}}\bar\mu}{3\pi |c|^3} + \frac{
  4(\bar\mu)^{\frac{5}{2}}+16(\tilde\mu)^{\frac{5}{2}} }{15\pi
  |c|^3},\nonumber\\
 Q^2&\approx &
  \tilde\mu-\frac{2p^b}{|c|}+\frac{64(\bar{\mu})^{\frac{3}{2}}\tilde\mu}{3\pi
  |c|^3}+ \frac{64(\bar\mu)^{\frac{5}{2}}}{15\pi |c|^3}.
\end{eqnarray}
After eliminating $B$ and $Q$, we have
\begin{eqnarray}
p^b &\approx& \frac{8}{3\pi}\left(\bar\mu-\frac{p^b+p^u}{2|c|}
+\frac{24(\bar\mu)^{\frac{5}{2}}+
16(\tilde\mu)^{\frac{5}{2}}+80(\tilde\mu)^{\frac{3}{2}}\bar\mu }{15\pi
|c|^3} \right)^{\frac{3}{2}}\nonumber\\ &
&-\frac{160(\bar\mu)^{3}+640(\bar\mu)^{\frac{3}{2}}(\tilde\mu)^{\frac{3}{2}}
}{45\pi^2|c|^3},\\ p^u& \approx & \frac{2}{3\pi}\left(
\tilde\mu-\frac{2p^b}{|c|}+\frac{64(\bar\mu)^{\frac{3}{2}} \tilde\mu
}{3\pi |c|^3} +\frac{64(\bar\mu)^{\frac{5}{2}} } {15\pi |c|^3}
\right)^{\frac{3}{2}} -\frac{128
(\bar\mu)^{\frac{3}{2}}(\tilde\mu)^{\frac{3}{2}} } {9\pi |c|^3}.
\end{eqnarray}

Obviously, the pressures $p^b$ and $p^u$ are functions of
$\bar{\mu}$, $\tilde{\mu}$ and the interaction strength $c$, i.e., 
$p^b=p^b(\bar\mu,\tilde\mu,|c|)$ and $p^u=p^u(\bar\mu,\tilde\mu,|c|)$.
Furthermore,  taking into account the relations  $\frac{\partial
p^b}{\partial H}+ \frac{\partial p^u}{\partial H}={P}/{2}$ and 
$\frac{\partial p^b}{\partial \mu}+ \frac{\partial p^u}{\partial
\mu}=n$, after a tedious calculation we find the effective
chemical potentials for the pairs $\mu^b=\mu+\epsilon_b/2$ and for the
unpaired fermions  $\mu^u=\tilde{\mu}=\mu+H/2$.
Explicitly,
\begin{eqnarray}
\mu^{u}&\approx &
\frac{\hbar^2n^2\pi^2}{2m}\left\{P^2+\frac{(1-P)(49P^2-2P+1)}{12|\gamma|}\right.\nonumber\\
& &\left. +\frac{(1-P)^2(93P^2+2P+1)}{8\gamma^2}-\frac{(1-P)}{240|\gamma|^3}\left[1441\pi^{2}P^{4}-7950P^4\right.\right.\nonumber\\
&
  &\left.\left. -324\pi^2P^3+15720P^3-7620P^2+166\pi^2P^2-120P-4\pi^2P-30+\pi^2\right]\right\},\label{mu-u}\\
\mu^{b} &\approx
&\frac{\hbar^2n^2\pi^2}{2m}\left\{\frac{(1-P)^2}{16}+\frac{(3P+1)(6P^2-3P+1)}{12|\gamma|}\right.\nonumber\\
&&\left.+\frac{(1-P)(5+17P-P^2+491P^3)}{64\gamma^2}+\frac{1}{240|\gamma|^3}\left[15(1+2P^2)+7470P^3+10\pi^2P^2\right.\right.\nonumber\\
& &\left.\left.-180\pi^2P^3+335\pi^2P^4-420\pi^2P^5-15405P^4-\pi^2+75P+7815P^5\right]\right\}.\label{mu-b}
\label{mu}
\end{eqnarray}

These results give
rise to a full characterization of two Fermi surfaces. The total
chemical potential can be determined from either
$\mu^b=\mu+\epsilon_b/2$ or  from  $\mu^u=\mu+H/2$. The chemical
potentials for the fermions with spin-up and spin-down states are
determined by $\mu_{\uparrow}=\mu+H/2$ and
$\mu_{\downarrow}=\mu-H/2$.
The energy for the model with arbitrary
population imbalances can be obtained from $E/L=\mu n-G +HP/2$, with result
\begin{eqnarray}
\frac{E}{L}&\approx&\frac{\hbar^2n^3\pi^2}{2m}\left\{-\frac{(1-P)\gamma^2}{4}
+\frac{\pi^2(1-3P+3P^2+15P^3)}{48}\right.\nonumber\\
&&\left.+\frac{\pi^2(1-P)(1+P-5P^2+67P^3)}{48|\gamma|}+\frac{\pi^2(1-P)^2(1+5P+3P^2+247P^3)}{64\gamma^2}\right.\nonumber\\
&&\left.-\frac{\pi^2(1-P)}{1440|\gamma|^3}\left[-15+31125{P}^{4}+1861{\pi }^{2}{P}^{5}-15765P^5-659{\pi }^{2}{P}^{4}\right.\right.\nonumber\\
& &\left.\left.+346{\pi }^{2}{P}^{3}-14{\pi }^{2}{P}^{2}+\pi^2
P+{\pi }^{2}-105P-150P^2-15090P^3\right]\right\}.\label{Energy}
\end{eqnarray}
This result provides higher order corrections in terms of the interaction
strength $\frac{1}{|\gamma|}$ compared to previous studies \cite{GBLB,Wadati}.

\section{Quantum Phase Transitions}
\label{diagram}

\begin{figure}
\vspace{0.5cm}
{{\includegraphics [width=0.80\linewidth]{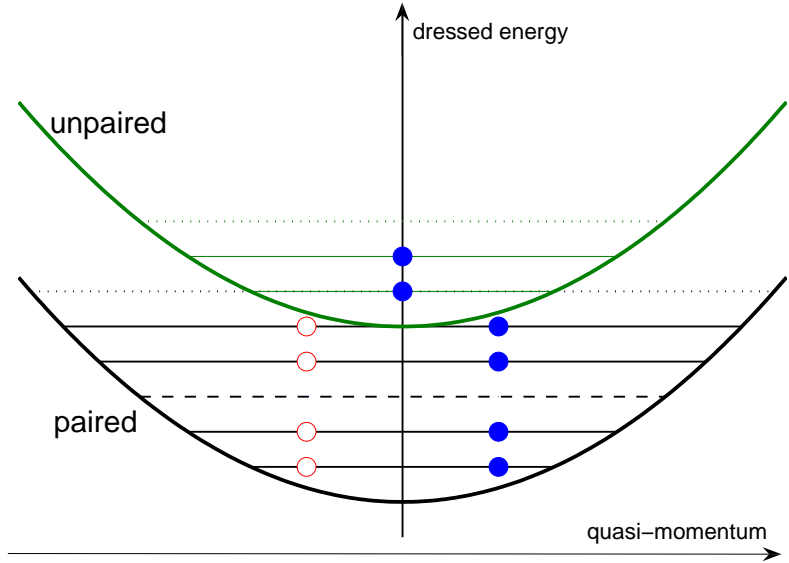}}}
\caption{Schematic dressed energy configuration for the gapless
phase in the vicinity of $H_{c1}$.}\label{fig:G-2}
\end{figure}

In section \ref{magnetization} we examined magnetic effects and
phase transitions for spin-1/2 weakly attractive fermions with polarization. 
As the attractive interaction strength $|\gamma|$ increases, the bound pairs become 
stable and form a singlet ground state. 
The ground state configuration is characterized by an empty unpaired Fermi sea, 
whereas the Fermi sea of the bound pairs is  filled up to the Fermi surface.  
The first critical field value $H_{c1}$ diminishes the gap, thus the excitations are gapless. 
This critical field indicates a phase transition from a
singlet ground state into a gapless phase where two Fermi liquids
of paired and unpaired fermions couple to each other. 
These configurations are depicted in Figure \ref{fig:G-2}.

Analysis of the dressed energy equations (\ref{TBA-F}) reveals that a fully
paired phase with magnetization $M^z=0$ is stable when the field $H<H_{c1}$, where
\begin{equation}
H_{c1}\approx \frac{\hbar^2n^2}{2m} \left[ \frac{\gamma^{2}}{2}
-\frac{\pi^2}{8} \left( 1-\frac{3}{4|\gamma|^2}-\frac{1}{|\gamma|^3}\right) \right].
\label{Hc1}
\end{equation}
In the vicinity of the critical field $H_{c1}$, the system exhibits a linear
field-dependent magnetization
\begin{equation}
M^z\approx\frac{2(H-H_{c1}) }{n {\pi }^{2}}
\left(1+ \frac{2}{|\gamma|}+  \frac{11}{2{\gamma}^2}+\frac{81-\pi^2}{6|\gamma|^3}\right)\label{Mz1}
\end{equation}
with a finite susceptibility
\begin{equation}
\chi\approx  \frac{2 }{n {\pi }^{2}}
\left(1+ \frac{2}{|\gamma|}+  \frac{11}{2{\gamma}^2}+\frac{81-\pi^2}{6|\gamma|^3}\right).\label{chi1}
\end{equation}

\begin{figure}[ht]
\vspace{0.5cm}
{{\includegraphics [width=0.60\linewidth,angle=-90]{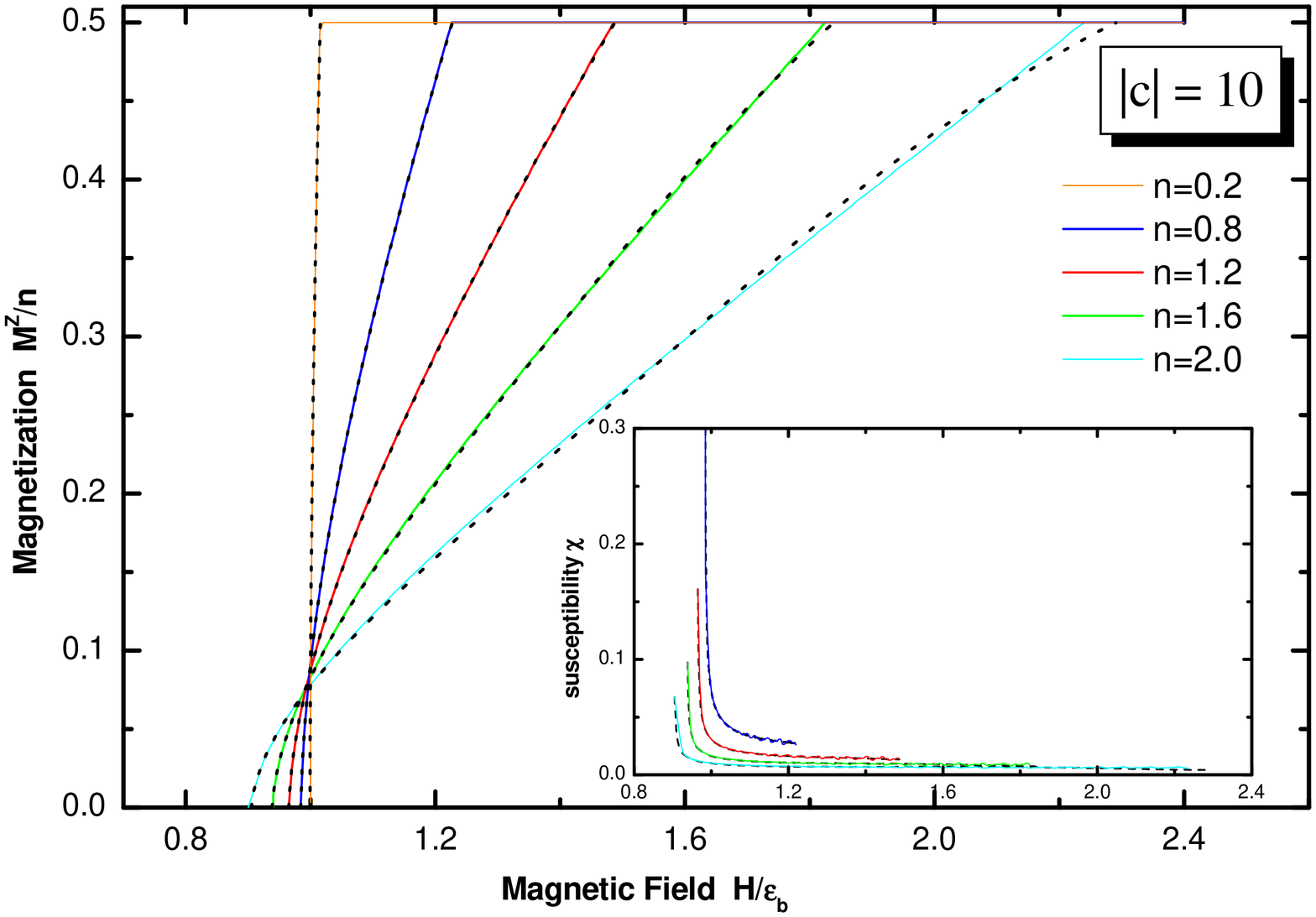}}}
\caption{Magnetization $M^z/n$ vs the external field $H/\epsilon_b$ for
$c=-10$ in the units $2m=\hbar=1$ for different densities $n$. The dashed lines are plotted
from the analytic result (\ref{E-Mz}). The solid curves are obtained from numerical solutions of 
the dressed energy equations (\ref{TBA-F}). 
Excellent agreement is seen between the analytic and numerical results. 
The inset shows similar comparison between analytic and numerical results for 
the susceptibility vs external field $H/\epsilon_b$.}\label{fig:mz-s}
\end{figure}

\begin{figure}[ht]
{{\includegraphics [width=0.80\linewidth]{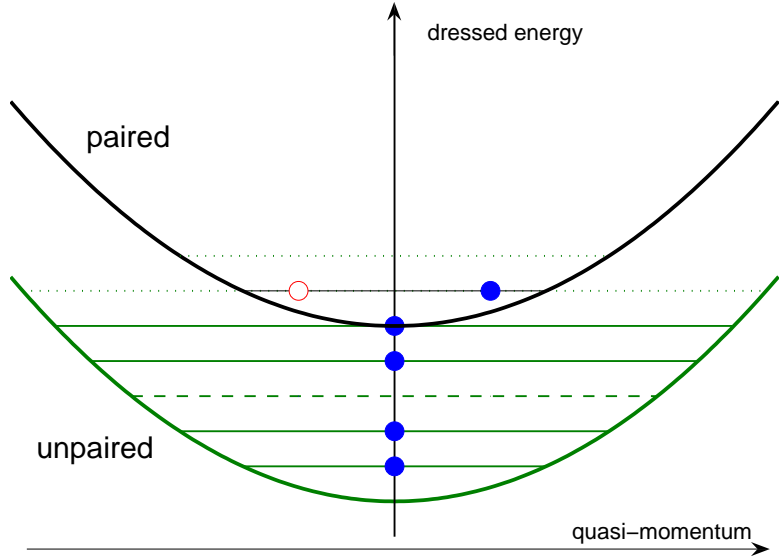}}}
\caption{Schematic dressed energy configuration for the gapless
phase in the vicinity of $H_{c2}$.}\label{fig:G-3}
\end{figure}

This universality class of linear field-dependent magnetization behaviour
 is also found for the multicomponent Fermi gases with
attractive interaction \cite{GBLZ}. 
However, it differs subtly from the case of a 
Fermi-Bose mixture due to the different statistical signature of 
a boson and a bound pair of fermions with opposite spin states \cite{GBL}.
For the model under consideration here the magnetic properties in this gapless phase can be exactly
described by the  external field-magnetization relation
\begin{equation}
\frac12{H}=\frac12{\epsilon_b}+\mu^u-\mu^b\label{E-Mz}
\end{equation}
where $\mu^u$ and $\mu^b$ are given by (\ref{mu-u}) and (\ref{mu-b}) with
$P=2M^z/n=2m^z$.  
This relation reveals an important
energy transfer relation among the binding energy, the variation of
Fermi surfaces and the external field.  
This relation might provide evidence
for the pairing signature in a 1D imbalanced Fermi gas with 
attractive interaction, i.e., pairs with nonzero centre-of-mass momenta. 
The lower critical field is reminiscent of the Meissner effect, whereas the
upper critical field determined by  (\ref{E-Mz}) is reminiscent of a
quantum phase transition from superconducting to normal states \cite{Clogston}.
Figure \ref{fig:mz-s} shows the magnetization vs external field for
different values of the interaction strength $\gamma$.
Numerical solution of the dressed energy equations (\ref{TBA-F}) 
shows that the analytic results are highly accurate in the 
strong and finitely strong coupling regimes.

A similar configuration occurs for the external field exceeding the upper
critical field $H_{c2}$, given by
\begin{equation}
H_{c2}\approx  \frac{\hbar^2n^2}{2m} \left[ \frac{{\gamma}^{2}}{2}+2{\pi
}^{2}\left( 1-\frac{4}{3|\gamma|}+\frac{16\pi^2}{15|\gamma|^3}\right) \right]
\label{Hc2}
\end{equation}
where a phase transition from the mixed phase into the normal  Fermi liquid phase occurs. 
Figure \ref{fig:G-3} shows this configuration in the dressed energy language.
{}From the relation (\ref{E-Mz}), we obtain the linear field-dependent magnetization as
\begin{equation}
M^z\approx\frac{n}{2}\left[ 1- \frac{H_{c2}-H}{4n^2\pi^2} \left( 1 + \frac
{4}{|\gamma|} +\frac{12}{{\gamma}^{2}} -\frac{16({\pi
}^{2}-6)}{3{|\gamma|}^{3}} \right)    \right] \label{Mz-2}
\end{equation}
with a finite susceptibility
\begin{equation}
\chi\approx  \frac {1}{8n{\pi }^{2}} \left(1 + \frac {4}{|\gamma|}
+\frac{12}{{\gamma}^{2}} -\frac{16({\pi }^{2}-6)}{3{|\gamma|}^{3}} \right).\label{chi2}
\end{equation}

\begin{figure}[ht]
{{\includegraphics [width=0.60\linewidth,angle=-90]{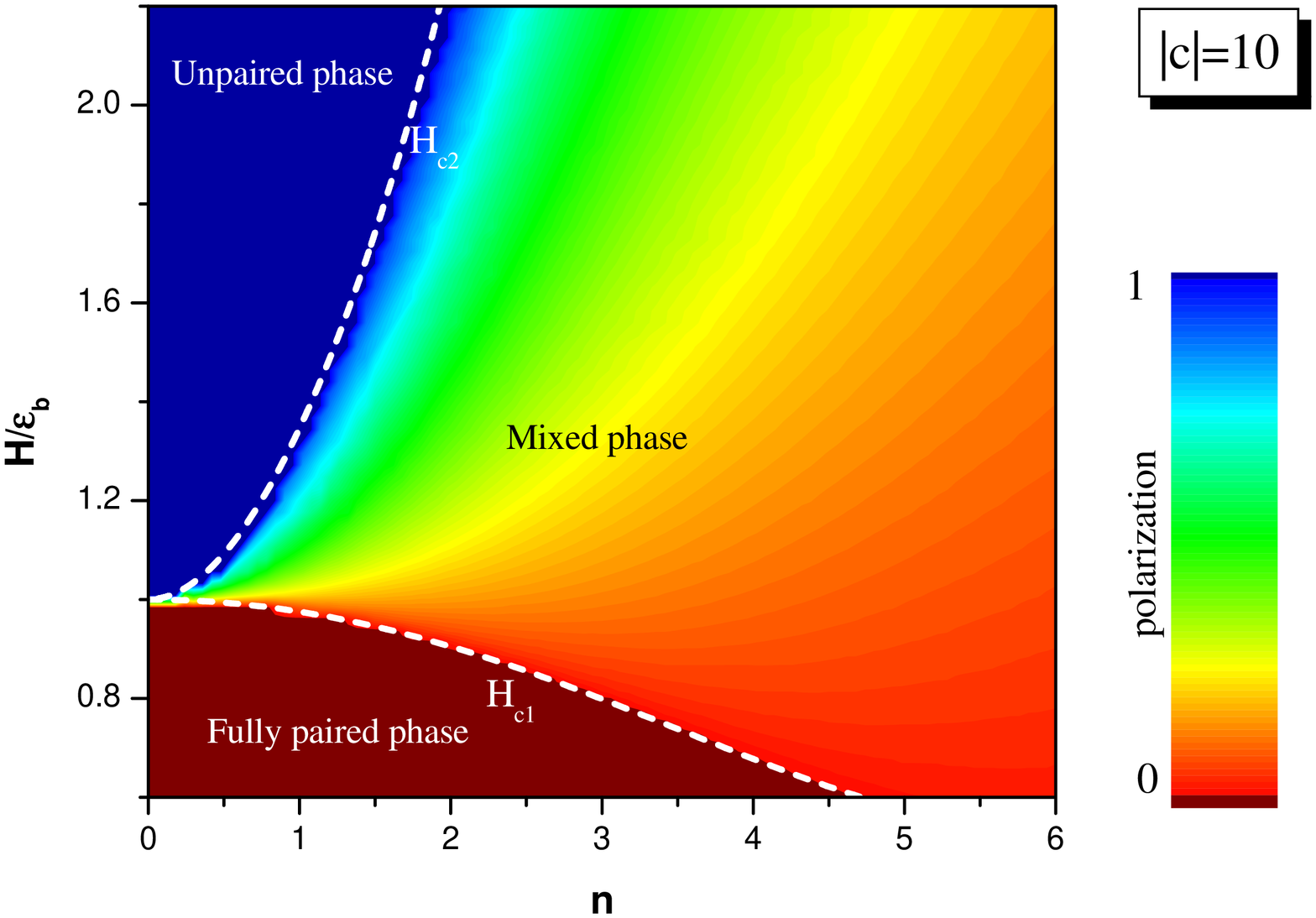}}}
\caption{Phase diagram for finite strong interaction with $|c| =10$. The
  dashed lines are plotted from Eqs. (\ref{Hc1}) and (\ref{Hc2}). 
  The coloured phases are obtained by numerical solution of the 
  dressed energy equations (\ref{TBA-F}). The numerical phase
  transition boundaries coincide well with the analytic results 
  (\ref{Hc1}) and (\ref{Hc2}). }\label{fig:phase-s}
\end{figure}

A typical phase diagram in the $n-H$ plane for
finite strong interaction is shown in Figure \ref{fig:phase-s}. 
Smooth magnetization curves at the critical fields $H_{c1}$ and $H_{c2}$ indicate second order phase
transitions.  
Very good agreement is observed between the curves obtained from the numerical solution of the
dressed energy equations and the analytical predictions (\ref{Hc1}) and (\ref{Hc2}) for the critical fields.

\section{Conclusion}
\label{Conclusion}

In summary, we have studied magnetic properties and quantum phase transitions 
for the 1D Bethe ansatz integrable model of spin-1/2 attractive fermions. 
Previous work on this model has been extended to derive higher order corrections 
for the ground state energy, pressure, chemical potentials,
magnetization, susceptibility and critical fields in
terms of the external magnetic field, density and interaction strength.
The range and applicability of the analytic results have been compared favourably with numerical solutions 
of the dressed energy equations.
The universality class of
linear field-dependent behaviour of the phase transitions in the
vicinity of the critical field values has been predicted for the whole attractive
regime. This universal behaviour is consistent with the prediction for the 1D
Hubbard model \cite{Penc}. However, it appears not to support the
argument \cite{Schlot,Schlot2} for a van Hove-type singularity of 
quantum phase transition for 1D attractive fermions.
Finite temperature properties of 1D interacting fermions will be considered elsewhere.

We further confirm that 1D strongly attractive fermions with population imbalance 
exhibit three quantum phases, subject to the value of the external field $H$ \cite{GBLB}: 
(i) for $H < H_{c1}$ bound pairs form a singlet ground state, 
(ii) for $H > H_{c2}$ a completely ferromagnetic phase without pairing occurs, and
(iii) for the intermediate range $H_{c1} < H < H_{c2}$ paired and unpaired atoms coexist.  
The typical phase diagram is as depicted in Figure \ref{fig:phase-s}.
However, for weak coupling, the BCS-like pairing is unstable.  
Two quantum phases emerge when the external field is applied: 
(i) a fully polarized phase for $H > H_{c}$, and (ii) a coexisting phase of paired and
unpaired fermions for $0<H<H_{c}$.  
The phase diagram for weak coupling is illustrated in Figure \ref{fig:phase-w}.
We have shown that the
mixed phase in 1D interacting fermions with polarization can be
effectively described by two coupled Fermi liquids.  Our exact phase
diagrams for the weak and strong coupling regimes also provide a space
segment signature for an harmonically trapped Fermi gas in 1D geometry.
These quantum phases and magnetic properties may also possibly
be observed in experiments with ultracold fermionic atoms \cite{Jin,Ketterle3}.

\vskip 5mm
\noindent
{\bf Acknowledgements}

This work has been supported by the Australian Research Council.  The
authors thank Miguel Cazalilla, Erhai Zhao and Chaohong Lee for
helpful discussions.  The author JSH is supported by the
NSFC (10671187), he also thanks the Department of Theoretical Physics,
Australian National University for their
hospitality during his visit in Dec 2007. AF thanks CNPq (Conselho
Nacional de Desenvolvimento Cientifico e Tecnologico) for financial
support.  XWG thanks the Department of Mathematics, University of Science
and Technology of China and MTB thanks the Centre for Modern Physics, 
Chongqing University, China for hospitality during various stages of this work.


\end{document}